\documentclass[12pt, onecolumn, a4paper, journal]{IEEEtran}
\usepackage{multirow}
\usepackage[dvips]{graphicx}
\usepackage{amsmath}
\usepackage[psamsfonts]{amssymb}
\usepackage{amsxtra}
\usepackage{threeparttable}
\usepackage{graphicx}
\usepackage{amsmath}
\usepackage{epsfig}
\usepackage{indentfirst}
\usepackage{float}
\usepackage{amssymb}
\usepackage{color}

\begin{document}

\title{Generalized Cross-correlation Properties of Chu Sequences}
\author{Jae Won Kang,
Younghoon Whang, Hyo Yol Park and \\ Kwang Soon
Kim$^{\dagger}$,~\IEEEmembership{Senior
Member,~IEEE}\\[2mm]
\authorblockA{
Department of Electrical and Electronic Engineering,\\
 Yonsei University, \\
 134 Shinchon-dong, Seodaemun-gu, Seoul 120-749, Korea\\
 Tel: +82-2-2123-5861\\
 Fax: +82-2-313-2879\\
 E-mail: ks.kim@yonsei.ac.kr\\[2mm]
$^{\dagger}$
 Corresponding Author\\ [2mm]} }% <-this % stops a space
\markboth{Manuscript Submitted to IEEE Transactions on Information
Theory} \small\normalsize \maketitle \def\baselinestretch{1.5}
\begin{abstract}
In this paper, we analyze the cross-correlation properties for Chu
sequences, which provide information on the distribution of the
maximum magnitudes of the cross-correlation function. Furthermore,
we can obtain the number of available sequences for a given
maximum magnitude of the cross-correlation function and the
sequence length.
\end{abstract}
\begin{IEEEkeywords}
Chu sequences, cross-correlation function.
\end{IEEEkeywords}
\newpage
\IEEEpeerreviewmaketitle
\def\baselinestretch{2.0}
\small\normalsize
\section{Introduction}
\IEEEPARstart{I}{n} general, it is desired to design a set of
sequences  with an impulsive autocorrelation function and a zero
cross-correlation function for many practical applications.
However, according to the Welch bound, the Sarwate bound, the
Sidelnikov bound, the Massey bound and other bounds
\cite{sarwate}--\cite{massey}, it was shown to be impossible to
construct such an ideal set of sequences. Therefore, searching
large families of sequences with good auto-correlation function
and cross-correlation function properties has been one of the most
interesting topics in sequence design. For evaluating the
correlation properties, one good choice is to use the maximum
sidelobe magnitude of the autocorrelation function and the maximum
magnitude of the cross-correlation function, which are
respectively denoted as $\hat \theta _a$ and $\hat \theta _c$ in
this paper. Here, the following questions arise naturally: how
many pairs of sequences are available for a given maximum values
of $ \hat \theta _a$ and $\hat \theta _c$ and what is the
distribution of the magnitude of the cross-correlation
function?\\
\indent Among well known good sequences are Kasami \cite{kasami},
Gold \cite{kasami}, Chu \cite{zadoff}--\cite{frank} and complex
four-phase \cite{4phase} sequences. For Kasami and Gold sequences,
it was shown that there are $\sqrt{N+1}$ sequences satisfying $
\hat \theta _a=1$ and $ \hat \theta _c=1+\sqrt{2/N}$
\cite{kasami}\cite{gold}\cite{sarwate2}\cite{sipark}, where $N$ is
the sequence length. For four-phase sequences, the number of
sequences satisfying $\hat \theta _a=1+\sqrt{N}$ and $ \hat \theta
_c=1+\sqrt{N}$ is $N+2$ \cite{4phase}\cite{fan}\cite{srpark}. On
the other hand, the autocorrelation function of Chu sequences is
known to be zero except at the lag of an integer
multiple of the sequence length \cite{zadoff}--\cite{frank}\cite{general}--\cite{exchu}.\\
\newtheorem{definition}{\textbf{Definition}}
\indent A set of Chu sequences with length \(N\) is defined as $C
= \left\{ {a_r \,|\,0 < r<N,\,\,\gcd (N,r) = 1} \right\}$, where
the $k$th element of $a_r$, $a_r(k)$, is defined as
\begin{equation} \label{eq1}
a_r(k)=\left\{ {\begin{split}
   &{\exp \left( {j\pi \frac{{rk^2 }}{N}} \right)\qquad \;\;,\,\,N\,\,\text{even},}  \\
   &{\exp \left( {j\pi \frac{{rk(k + 1)}}{N}} \right),\,\,N\,\,\text{odd}.}  \\
\end{split}} \right.
\end{equation} The periodic autocorrelation function with lag
$\tau$, $\theta_r(\tau)$, of the sequence $a_r$ is defined as
\begin{equation}\label{eq2}
\theta_r(\tau)  =  \sum\limits_{k = 0}^{N - \tau - 1} {a_r (k)a_r^
* (k + \tau) + } \sum\limits_{k = N - \tau}^{N - \tau - 1} {a_r
(k)a_r^* (k + \tau - N)}.
\end{equation}
In \cite{chu}, it was shown that the periodic autocorrelation
function of Chu sequences satisfies
\begin{equation}\label{eq3}
 \theta_r(\tau) = \left\{ {\begin{array}{*{20}c}
   {N,\,\,\tau\,\,\bmod \,N = 0},  \\
   {0,\,\,\,\tau\,\,\,\bmod \,N \ne 0}.  \\
\end{array}} \right.
\end{equation}
Let \(a_r\) and \(a_s\) be any two Chu sequences with length $N$.
Then, the cross-correlation function \(\theta_{r,s}(\tau)\) of
$a_r$ and $a_s$ with lag $\tau$ is defined as
\begin{equation}\label{eq4}
\begin{split}
 \theta_{r,s} (\tau)&= \sum\limits_{k = 0}^{N - \tau -
1}{a_r (k)a_s^
* (k + \tau)}+  \sum\limits_{k = N - \tau}^{N - \tau - 1} {a_r (k)a_s^ *  (k
+ \tau - N)}\\
 &= \sum\limits_{k = 0}^{N - 1} {a_r (k)a_s^ * (k + \tau)},
\end{split}
\end{equation}
where the last equality comes from the fact that $a_r (k+d)=a_r
(k+d+N)$ for an arbitrary integer $d$ \cite{chu}.\\
\indent In \cite{sarwate} and \cite{peng}, it was shown that the
maximum magnitude of the cross-correlation function $ \hat \theta
_c$ can be lower-bounded as a function of the sequence length and
the maximum magnitude of the autocorrelation function, $ \hat
\theta _a$. By using this lower-bound, the optimum correlation
properties of a set of sequences can be defined and it follows
that the lower bound of $ \hat \theta _c$ is equal to \(\sqrt{N}\)
when $ \hat \theta _a$ equals zero. Certain pairs of Chu
sequences, $a_r$ and $a_s$, meet this lower-bound when $\gcd(r-s,
N)=1$. However, in order to obtain more Chu sequences with
relatively low cross-correlation values, we need to investigate
more general cross-correlation
properties.\\
\indent In this paper, we derive general properties for
cross-correlation function of Chu sequences. Using the derived
properties, we can obtain the magnitude distribution of the
cross-correlation function. Here, the maximum magnitude denotes
the maximum magnitude value of the cross-correlation function of
two given Chu sequences among all possible lags and its
distribution is taken over all possible pairs of Chu sequences. In
addition, the number of available sequences can be obtained for a
given value of $\hat \theta _c$ and the given sequence length.\\
The remaining of this paper is organized as follows. In Section
II, the magnitude of cross-correlation function of Chu sequences
are described. In Section III, the distribution of the maximum
magnitude of the cross-correlation function and the number of
available Chu sequences for given maximum cross-correlation value
and the sequence length are investigated. Finally, Section IV
concludes this paper.
\section{Characteristic of the Cross-correlation function of Chu sequences}

\newtheorem{theorem}{\textbf{Theorem}}
In order to investigate the cross-correlation function of Chu
sequences in detail, we need to find what are the possible values
that the cross-correlation function of Chu sequences can take,
which are given in the following theorem.\\
\begin{definition}\label{df1}
 Let $r$ and $s$ be positive integers satisfying $0<r,s<N$,
$\gcd (N,r)=1$ and $\gcd (N,s)=1$. Also define $g_{r,s}=\gcd
(N,r-s),\, u_{r,s}=N/g_{r,s}$ and $v_{r,s}=(r-s)/g_{r,s}$. Then
$u_{r,s}$ is relatively prime with $v_{r,s}$. Also, for a given
lag $\tau$, we can rewrite it as $\tau=i_\tau g_{r,s} + d_\tau$,
where $i_{\tau} =\left\lfloor {{\tau}/g_{r,s}} \right\rfloor$ and
$d_\tau=
\tau- i_{\tau}g_{r,s}$.\\
\end{definition}
\begin{theorem}\label{theorem1}
The magnitude of the cross-correlation function $\theta_{r,s}
(\tau)$, $\left| {\theta_{r,s} (\tau)} \right|$, is given as
\[
\begin{array}{l}
\left| {\theta_{r,s} (\tau)} \right|  =
 \left\{ {\begin{split}
    &\sqrt{Ng_{r,s}} \delta _K \left( { d_\tau } \right),\,\,\,\,\,\,\,\,N\,\,\, \text{and}\,\,u_{r,s}v_{r,s} \,\,\text{even},\,\,\text{or}\,\, N\,\,\text{odd},   \\
    &\sqrt{Ng_{r,s}} \delta _K \!\left( {d_\tau-\frac{g_{r,s}}{2}} \right),\,\,\,\,\,\,\,\,N\,\,\text{even}\,\,\text{and}\,\,u_{r,s} v_{r,s} \,\,\text{odd},  \\
   &{\,\,\,\,\,\,\,\,0,\,\,\,\,\,\,\,\, \text{otherwise}},  \\
\end{split}} \right.
\end{array}
\]
where $\delta _K \left( {\cdot} \right)$ is the Kroneker delta
function.\\
\end{theorem}
\indent To prove Theorem \ref{theorem1}, the following lemmas are useful.\\\\
 \indent\textit{\textbf{Lemma }1} \cite{zadoff}\cite{chu}\textit{:} \label{Lemma1} The $h$th primitive
root of unity $\xi_{h}$ can be defined as $\xi_{h} = \exp \left(
{j2\pi \frac{u}{h}} \right)$, where $u$ is any integer relatively
prime to $h$. Then, for any integer $v$, $0<v\leq h-1$,
\[
\sum\limits_{k = 0}^{h - 1} {\xi_{h}^{ \pm vk}
}=0,\,\,\,\,\,\,\,\xi_{h}  \ne 1.
\]\\
 \indent\textit{\textbf{Lemma }2}\textit{:} The squared magnitude
 of the cross-correlation function is given as
\[
\begin{array}{l}
 \left| {\theta _{r,s} (\tau)} \right|^2=  \\
 \left\{\begin{split}
  {u_{r,s}g_{r,s} \sum\limits_{m = 0}^{g_{r,s}  - 1} {( - 1)^{u_{r,s} v_{r,s} m^2 } \exp\left( {j2\pi \frac{{smd}}
{{g_{r,s} }}} \right)\!,\,N\,\text{even}},}\\
   {u_{r,s} g_{r,s} \sum\limits_{m = 0}^{g_{r,s}  - 1} {( - 1)^{v_{r} m(u_{r,s}  + 1)} \exp \left( {j2\pi \frac{{smd}}
{{g_{r,s} }}} \right),\,N\,\text{odd}}.}\\
\end{split}  \right.
\end{array}
\]
The proof of Lemma 2 is given in Appendix A.\\\\
\indent Now, the proof of Theorem \ref{theorem1} is given as
follows.
\begin{IEEEproof}[\text{Proof of Theorem \ref{theorem1}}]
First, consider the case when $N$ and $u_{r,s}v_{r,s}$ are even.
Then, from Lemma 2, we obtain
\begin{equation}\label{eq5}
\left| {\theta_{r,s} (\tau)} \right|^2 =u_{r,s}
g_{r,s}\sum\limits_{m = 0}^{g_{r,s} - 1} {\exp \left( {j2\pi
\frac{{smd_\tau}}{g_{r,s}}} \right)}.
\end{equation}
If $d_\tau=0$, $\left|\theta_{r,s}(i_\tau u_{r,s})
\right|^2=u_{r,s} g_{r,s}^2=Ng_{r,s}$. When, $d_{\tau} \neq 0$,
since $s$ is
relatively prime with $g_{r,s}$, $\left|\theta_{r,s}(i_\tau g_{r,s}+d_\tau)\right|^2=0$.\\
\indent Now, consider the case when $N$ is even and $u_{r,s}
v_{r,s}$ is odd. When $d_\tau=g_{r,s}/2$, we obtain from Lemma 2
that
\begin{equation}\label{eq6}
\begin{split}
 \left|\theta_{r,s}  \left( {i_\tau g_{r,s}+ \frac{g_{r,s}}{2}} \right)\right|^2=u_{r,s} g_{r,s}\sum\limits_{m = 0}^{g_{r,s} - 1}  {\exp \left\{
{j2\pi  \left(
{ - \frac{{m(u_{r,s} v_{r,s} m + s)}}{2}} \right)} \right\}}.\\
\end{split}
\end{equation}
We know that $s$ is odd because $s$ is relatively prime with $N$.
If $m$ is odd, $u_{r,s} v_{r,s} m$ is odd and $u_{r,s} v_{r,s}
m+s$ is even. On the other hand, if $m$ is even, $u_{r,s} v_{r,s}
m$ is also even. Thus, $m(u_{r,s} v_{r,s} m + s)$ is always even
and it shows that $\left|\theta_{r,s} (i_\tau g_{r,s} +
g_{r,s}/2)\right|^2 =u_{r,s} g_{r,s}^2=Ng_{r,s}$. When $d_\tau
\neq g_{r,s}/2$, from Lemma 2, we can rewrite
$\left|\theta_{r,s}(\tau)\right|^2$ as
\begin{equation}\label{eq7}
\begin{split}
\left|\theta_{r,s} {(i_\tau g_{r,s} + \frac{g_{r,s}}{2}+d_\tau
')}\right|^2&= u_{r,s} g_{r,s}\sum\limits_{m = 0}^{g_{r,s} - 1}
{\exp } \left\{ {j2\pi \left( { - \frac{{m(u_{r,s}v_{r,s} m +
s)}}{2}
+ \frac{{smd_\tau ' }}{g_{r,s}}} \right)} \right\}\\
&=u_{r,s} g_{r,s}\sum\limits_{m = 0}^{g_{r,s} - 1} {\exp }
\left( {j2\pi \frac{{smd_\tau ' }}{g_{r,s}}} \right)\\
&=0,
\end{split}
\end{equation}
where $d'_\tau=d_\tau-g_{r,s}/2$ and the last equality comes from
the fact that $s$ is relatively prime with $g_{r,s}$. \\
\indent Finally, consider the case where $N$ is odd. Then
$g_{r,s}$ and $u_{r,s}$ should be odd. Then, from Lemma 2, we
obtain
\begin{equation}\label{eq8}
\begin{split}
\left|\theta_{r,s} {(i_\tau g_{r,s}+d_\tau )}\right|^2 &=u_{r,s}
g_{r,s}\sum\limits_{m = 0}^{g_{r,s} - 1} {\exp \left( {j2\pi
\frac{{smd_\tau}}{g_{r,s}}} \right)}\\
&=\sqrt{Ng_{r,s}} \delta _k \left( { d_\tau } \right),
\end{split}
\end{equation}
which concludes the proof.
\end{IEEEproof}
\section{Distribution of the maximum magnitudes of the cross-correlation function}
\subsection{The uniform property}
Theorem \ref{theorem1} tells us that the characteristic of the
cross-correlation function of two Chu sequences, $a_r$ and $a_s$,
depends only on $g_{r,s}=\gcd(r-s,N)$. For example, when
$g_{r,s}=1$, $\hat \theta _c$ meets the lower bound of $\sqrt{N}$.
On the other hand, when $g_{r,s}=N$, $\hat \theta _c$ becomes the
largest value of $N$. However, it has not yet been investigated
how many sequences are available for a given values of $ \hat
\theta _c$ and the sequence length. To answer the question, it is
required to investigate the distribution of the maximum magnitude
values of the cross-correlation function.\\
\begin{definition}\label{df2}
Any given integer $N$ can be represented as $N=\prod\nolimits_{i =
1}^k {p_i^{c_i } }$, where $p_i$ denotes the $i$th smallest prime
factor of $N$. Let us define $ \mu _N  = \left\{ {n|0 < n <
N,\,\gcd (n,N) = 1} \right\}$ as the index set of Chu sequences of
length $N$. Also, for a given integer $c$, define the following
sets and function as follows.
\begin{itemize}
    \item $U_{N,c} = \{n-c\,|\, 0\leq n<N\}$
    \item $R_{N,c} = \{n-c\,|\,n \in \mu _N \}$
    \item $D_{N,c} = \{n-c\,|\, 0\leq n<N$ and $n \notin \mu_N \}$
    \item $P_{N,c}^m=\{np_m-c\,|\,0 \leq n < N/p_m\, \}$
    \item $G_{N,x}(S)=\{n\,|\,n \in   S$ and $\gcd(n,N)=x\}$
     for a given integer set $S$.
    \item $|A| :$ The cardinality of a set $|A|$.
\end{itemize}
\end{definition}
\indent From Theorem \ref{theorem1}, we can see
that the maximum magnitude of the cross-correlation function
between $a_r$ and $a_s$ is $ \hat \theta _{r,s}=\mathop {\max
}\limits_\tau  |\theta _{r,s} (\tau )| = \sqrt {g_{r,s} N}$. Thus,
for given $N$, $s \in \mu_N$ and $x$, it is easily seen that
$G_{N,x}(R_{N,s})=\{r-s\,|\,g_{r,s}=\gcd(r-s,N)=x$ and $r \in \mu
_N\}$ is the set of differences between $s$ and all Chu sequence
indices whose maximum squared magnitude of the cross-correlation
function with $a_s$ is equal to $x$. Then, $|G_{N,x}(R_{N,s})|$ is
the number of available Chu sequences satisfying $ \hat \theta
_{r,s}  ^2 / N =x$. Then, the main result of this subsection is
given in the following theorem.\\
\begin{theorem}\label{theorem2}
Let $1 \leq s \neq s' \leq N$ be two different integers relatively
prime with $N$. Then,
$|G_{N,x}(R_{N,s})|=|G_{N,x}(R_{N,s'})|$.\\
\end{theorem}
\indent Theorem \ref{theorem2} indicates that the distribution of
the maximum magnitudes of the cross-correlation function for a
given Chu sequence set can be obtained by fixing one sequence
arbitrarily and examining the cross-correlation functions with the
other sequences. The
following Lemmas 3--5 are useful the proof of Theorem \ref{theorem2}.\\\\
\indent \textit{\textbf{Lemma }3}\textit{:} For any two different
integers $c$ and $c'$, $G_{N,x}(U_{N,c})=G_{N,x}(U_{N,c'})$.\\
\begin{IEEEproof}
It has been proved that $\gcd (c  + mN ,N ) =\gcd (c,N)$
\cite{math1}\cite{math2}. Then, it is easily seen that $\{\,
\gcd(-c+1, N),$ $\gcd$$(-c+2,N) , \,\cdots,\,\gcd(-c+N,N) \,\} =
\{\, \gcd(1,N),$ $\gcd(2,N),\,\cdots ,\,\gcd(N,N)\,\}$ for any
integer $c$. Therefore, $G_{N,x}(U_{N,c})=$ $G_{N,x}(U_{N,c'})$.
\end{IEEEproof}
\vskip 24pt
\indent \textit{\textbf{Lemma }4}\textit{:} Let $a$
and $b$ be positive integers satisfying $\gcd(a,b)=1$. Also, for
an arbitrary positive integer $m$, define $C=\{na-c\,|\,k \leq n <
k'\}$, where $k$ is an arbitrary integer and $k'=k+mb$. Then, $C$
contains
exactly $m$ integer multiples of $b$.\\
\begin{IEEEproof}
For the $i$th element $c_i=k_i a-c$ of $C$, we can represent it as
$c_i=q(k_i)b+e(k_i)$, where $q(k_i)=\left\lfloor
{c_i/b}\right\rfloor$ and $e(b_i)=c_i$ mod $b$. Note that such a
pair of $q(k)$ and $e(k)$ is unique for a given $c_i$
\cite{math2}. Let $d_{ij}=c_i-c_j$. Then
$d_{ij}=(k_i-k_j)a=\{q(k_i)-q(k_j)\}b+{e(k_i)-e(k_j)}$. Thus,
$e(k_i)=e(k_j)$ implies that $(k_i-k_j)$ is an integer multiple of
$b$ and vice versa because $a$ is relatively prime with $b$. Now,
consider the partition $\{C_r,\,\,0\leq r<m\}$, where
$C_r=\{k_{r,i}a+c\,|\,k+rb \leq k_{r,i} < k+(r+1)b\}$. Then, each
$C_r$ contain exactly one element that is an integer multiple of
$b$ since $e(k_{r,i})$, $k+rb \leq k_{r,i} < k+(r+1)b$, are all
distinct and $0 \leq e(k_{r,i}) < b$, which concludes the proof.
\end{IEEEproof}
\vskip 24pt
\indent \textit{\textbf{Lemma }5}\textit{:} Let $1
\leq s \neq s' < N$ be two different integers relatively prime
with $N$. Then,
if $x$ is a divisor of $N$, $|G_{N,x}(D_{N,s})|=|G_{N,x}(D_{N,s'})|$.\\
\begin{IEEEproof}
For a given $N=\prod\nolimits_{i = 1}^k {p_i^{c_i } }$, since
$D_{N,s}=\bigcup\nolimits_{i = 1}^k {P_{N,s}^i }$ from Definition
2, $| {G_{N,x} \left( {D_{N,s} } \right)} | = \left| {G_{N,x}
\left( {\bigcup\nolimits_{i = 1}^k {P_{N,s}^i } } \right)}
\right|$. Then, $\left| {G_{N,x} \left( {D_{N,s} } \right)}
\right|$ can be rewritten as
\begin{equation}\label{eq9}
\begin{split}
&\left| {G_{N,x} \left( {D_{N,s} } \right)} \right|
=\sum\limits_{i = 1}^k {\left| {G_{N,x} \left( {P_{N,s}^i }
\right)} \right|}-\sum\limits_{i_1  = 1}^{k - 1} {\sum\limits_{i_2
= i_1 + 1}^k \!\!{\left| {G_{N,x} \left( {P_{N,s}^{i_1 }  \cap
P_{N,s}^{i_2 } } \right)} \right|}}+\cdots\\
&+ ( - 1)^{k - 2} \sum\limits_{i_1  = 1}^2 {\sum\limits_{i_2  =
i_1  + 1}^3 { \cdots \sum\limits_{i_{k - 1}  = i_{k - 2}  + 1}^k
{\left| {G_{N,x} \left( {\bigcap\limits_{m = 1}^{k - 1}
{P_{N,s}^{i_m } } } \right)} \right|}  } }+ \left( { - 1}
\right)^{k - 1} \left| {G_{N,x} \left( {\bigcap\limits_{i = 1}^k
{P_{N,s}^i } } \right)} \right|.
\end{split}
\end{equation}
\indent If $\gcd(x,p_m)=1$, it is seen easily that there always
exist $N/(p_m x)$ integer multiples of $x$ among the elements in
$P_{N,s}^{m}$ from Lemma 4. If $\gcd(x,p_m)\neq 1$, $p_m$ should
be a divisor of $x$ since $p_m$ is a prime number. Thus, there is
no integer multiple of $x$ among the elements in $P_{N,s}^{m}$
since $x$ is relatively prime with $s$. Thus,
$G_{N,x}(P_{N,s}^{m})$ does not depend on $s$ as long as $s$ is
relatively prime with $N$, i.e.,
$G_{N,x}(P_{N,s}^{m})=G_{N,x}\left( {P_{N,s'}^{m}} \right).$
\indent Let $M$ be an arbitrary subset of $\{1,\cdots,k\}$. Also,
let $m_i$ denote the $i$th element of $M$. Then, for a given index
set $M$, define $L_{N,s}^M=\bigcap\nolimits_{i = 1}^{|M|}
{P_{N,s}^{m_i} }=\left\{ {nl_m-s\,|\,0 \leq n < N/l_m} \right\}$,
where $l_m=\prod\nolimits_{i = 1}^{|M|} {p_{m_i } }$. Similarly,
if $\gcd(x,l_m)=1$, there always exist $N/\left( {l_{m}x} \right)$
integer multiples of $x$ among elements in $L_{N,s}^{M}$ from
Lemma 4. If $\gcd(x,l_m)\neq 1$, $x$ should be an integer multiple
of $p_{m_i}$ for some $m_i \in M$. Since $s$ is relatively prime
with all $p_{m_i}$, $m_i \in M$, there is no integer multiple of
$x$ among the elements in $L^M_{N,s}$. Thus,
$G_{N,x}(L_{N,s}^{M})$ does not depend on $s$ as long as $s$ is
relatively prime with $N$. Thus, from (\ref{eq9}),
$|G_{N,x}(D_{N,s})|=|G_{N,x}(D_{N,s'})|$.
\end{IEEEproof}
\vskip 24pt The proof of Theorem \ref{theorem2} is now given as
follows.
\begin{IEEEproof}[\text{Proof of Theorem \ref{theorem2}}]
When $x$ is not a divisor of $N$, $G_{N,x}(R_{N,s})=\emptyset$
regardless of $s$. When $x$ is a divisor of $N$, from Lemmas 3 and
5, we have already seen that $|G_{N,x}(U_{N,s})|$ and
$|G_{N,x}(D_{N,s})|$ does not depend on $s$ as long as $s$ is
relatively prime with $N$ and $x$ is a divisor of $N$. Also, from
Definition 2, since $R_{N,s}\cap D_{N,s}= \emptyset$, $\left|
G_{N,x} \left( {R_{N,s} } \right)\right|=\left|G_{N,x} \left(
{U_{N,s} } \right)\right|-\left| G_{N,x} \left( {D_{N,s} }
\right)\right|$. Thus, $G_{N,x}(R_{N,s})$ does not depend on $s$
as long as $s$ is relatively prime with $N$.
\end{IEEEproof}
\subsection{The distribution}
In this subsection, the distribution of the maximum magnitudes of
the cross-correlation function is investigated. The main result of
this subsection is given as follows.\\
\begin{theorem}\label{theorem3}
For $N=\prod\nolimits_{i = 1}^k {p_i^{c_i } }$, $0<x \leq N$, any
$s$ relatively prime with $N$, $|G_{N,x}(R_{N,s})|$ is given as
\begin{equation}\label{eq33}
|G_{N,x}(R_{N,s})|={ {\prod\limits_{i\in M_{x}}{\phi_x
(i)\prod\limits_{j \in M_{x}^k}{\Phi_x (j)}}}},
\end{equation}
where $M_{x}$, $M_{x}^k$, $\phi_x (i)$ and $\Phi_x (i)$ are
defined as follows. When $x$ is a divisor of $N$, $M_{x}$ denotes
the set of indices of the prime factors of $x$ and
$M_{x}^k=\{1,\cdots,k\}-M_x$. Then, $x$ can be represented as
$x=\prod\nolimits_{i \in M_{x}}{p_i^{n_i (x) }}$, where $0<n_i (x)
\leq c_i$. Also, $\phi_x (i)$ and $\Phi_x (i)$ are defined as
\[
\begin{split}
&\phi_x(i) \triangleq
   {p_i^{c_i  - n_i (x) }  -\left( {1 - \delta _K \left( {c_i  - n_i (x)} \right)} \right)p_i^{c_i  - n_i(x) - 1}},\\
&   \Phi_x(j)  \triangleq p_{j }^{c_{j }}  - 2p_{j}^{c_{j}  -1}.\\
\end{split}
\]
When $x$ is not a divisor of $N$, $M_{x} \triangleq \emptyset$,
$\phi_x(i)=0$ and $\Phi_x(i)=0$, so that $|G_{N,x}(R_{N,s})|=0$. \\
\end{theorem}
The following Lemmas 6--10 are useful for the proof of Theorem
\ref{theorem3}.\\\\
\indent \textit{\textbf{Lemma }6} (Euler function
\cite{math2})\textit{:} Let $\varphi (N)$ be the number of
positive integers that are relatively prime with
$N=\prod\nolimits_{i = 1}^k {p_i^{c_i } }$ among $\{n\,|\, 0< n <
N\}$. Then, $\varphi (N)$ is given as
\begin{equation}\label{eq10}
\varphi (N) = N\prod\limits_{i = 1}^k {\left( {1 - p_{i}^{ - 1}
}\right)}.
\end{equation}
\\
\indent \textit{\textbf{Lemma }7}\textit{:} If $x$ is not a
divisor of $N$, $|G_{N,x}(R_{N,s})|=0$ since $\gcd(r-s,N)$ cannot
be
equal to $x$.\\\\
 \indent \textit{\textbf{Lemma }8}\textit{:} When $N$ is a prime
 number, for $s \in \mu_N$ and $0<x\leq N$, $|G_{N,x}(R_{N,s})|$ is given as
\begin{equation}\label{eq11}
\begin{split}
|G_{N,x}(R_{N,s})| = \left\{ {\begin{array}{*{20}c}
   &1, &&\text{if}\,\, x=N , \\
   &N - 2,  &&\text{if}\,\, x=1,\\
   &0, &&\text{otherwise}.
\end{array}}\right.\\\\
\end{split}
\end{equation}
\begin{IEEEproof}
Since $N$ is a prime number, $\gcd(r-s,N)=1$ when $s \neq r$ and
$\gcd(r-s,N)=N$ when $s = r$. Thus, $|G_{N,x}(R_{N,s})|=0$ for
$1<x<N$. Since $\left| {\mu _N } \right| = \left| {R_{N,s} }
\right| = N - 1$ from Definition 2, $\left| {G_{N,1} \left(
{R_{N,s} } \right)} \right| = N - 2$ and $\left| {G_{N,N} \left(
{R_{N,s} } \right)} \right| = 1$ .
\end{IEEEproof}
\vskip 24pt
 \indent\textit{\textbf{Lemma }9}\textit{:} For
$N=\prod\nolimits_{i=1}^{k} {p_i}$, its divisor $x$ and
$\gcd(s,N)=1$, we can denote $M_x$ be the set of indices of the
prime factors of $x$ so that $x=\prod\nolimits_{i \in M_x}{p_i}$
and $M_{x}^k=\{1,\cdots,k\}-M_x$. Then, $|G_{N,x}(R_{N,s})|$ is
given as
\begin{equation}\label{eq12}
\begin{split}
\begin{array}{l}
 |G_{N,x}(R_{N,s})| =
 \left\{ {\begin{array}{*{20}c}
   1,\, &\text{if}\,\,x = N,  \\
\prod\limits_{i \in M_{x}^k} {\left( {p_{i } - 2}
\right)},\, &\text{if} \,\, x=\prod\limits_{i \in M_{x}}{p_{i } }\neq N,\\
0, &\text{otherwise}.
\end{array}} \right.
\end{array}
\end{split}
\end{equation}
\vskip 24pt
\begin{IEEEproof}
When $k=1$, (\ref{eq12}) holds from Lemma 8. Suppose
that (\ref{eq12}) holds for any divisor $x$ of $N$ when $k=K$ and
let $N'=Np_{K+1}$. Then, the set of all divisors of $N'$ is given
as $\{y=x\,\, \text{or}\,\, xp_{K+1} \,|\, x\in \{\text{all
divisors of } N\}\}$. Note that
$\mu_{N'}=\{\,n\,|\,0<n<N',\,\gcd(n,N')=1\}=\{\,n+mN\,|\,n\in
\mu_N,\, 0\leq m< p_{K+1},\, \gcd(n+mN,p_{K+1})=1\}$. Thus,
$R_{N',s}=\{\,n-s\,|\,n\in \mu_{N'}\,\}= \mathop  \bigcup
\nolimits_{m =0}^{p_{K + 1} - 1} R_{N,s}^m$, where $R_{N,s}^m  =
\{\, n + mN - s\,|\,n
\in \mu _N ,\,\gcd (n + mN,p_{K + 1} ) = 1\,\}$.\\
\indent When $y=x$ for a divisor $x$ of $N$, $G_{N',y} (R_{N,s}^m )
= \{ \,n + mN\,|\,n \in G_{N,x} (R_{N,s}),\,\gcd (n + mN,p_{K + 1} )
= 1,\,\gcd (n - s + mN,p_{K + 1} ) = 1\,\}$ since $\gcd(n-s+mN,
N')=x$ implies $\gcd(n-s,N)=x$ and $\gcd(n-s+mN,N')=x$ if
$\gcd(n-s,N)=x$ and $\gcd(n-s+mN,p_{K+1})=1$. Similarly, when
$y$=$xp_{K+1}$ for a divisor $x$ of $N$, $G_{N',y} (R_{N,s}^m ) = \{
\,n + mN\,|\,n \in G_{N,x} (R_{N,s} ),\,\gcd (n + mN,p_{K + 1} ) =
1,\,\gcd (n - s +
mN,p_{K + 1} ) = p_{K + 1} \,\}$. \\
\indent Now, define $B_{N} (n) = \{ \,n + mN\,|\,0 \leq m < p_{K +
1} \,\}$. Then, it is easily seen that
\begin{equation}\label{eq13}
\begin{split}
  |G_{N',y} (R_{N',s})| =& \sum\limits_{m = 0}^{p_{K + 1}  - 1} {|G_{N',y} (R_{N',s}^m )|}  \hfill \\
   =& \sum\limits_{n \in G_{N,x} (R_{N,s} )}{|B_{N} (n)| - } |\alpha _{N,N'}^s (x,y)|\hfill,
\end{split}
\end{equation}
where $\alpha _{N,N'}^s (x,y)= \{\, \zeta \,|\,\zeta\in \mathop
\bigcup \nolimits_{n \in G_{N,x} (R_{N,s}) } B_{N} (n),$ $\gcd
(\zeta+s ,$ $p_{K + 1} ) \ne 1\,\,\text{or}\,\,\gcd (\zeta ,p_{K+
1} ) \ne y/x\,\}$. From Lemma 4, it is easily seen that $|\alpha
_{N,N'}^s (x,y) \cap B_{N} (n)| = 2$ when $y=x$ since there is
exactly one element in $B_{N}(n)$ for each of the two conditions :
$\gcd(\zeta +s, p_{K+1}) \neq 1$ and $\gcd(\zeta, p_{K+1}) \neq 1
$ and the elements are different due to the fact that $\gcd
(\zeta+s ,p_{K + 1} ) \neq 1$ and $\gcd(s,N)=1$ implies $\gcd
(\zeta ,p_{K + 1} ) = 1$. Similarly, $p_{K+1}-1$ elements are not
integer multiples of $p_{K+1}$ in $B_{N} (n)$ when $y=xp_{K+1}$
since the first condition $\gcd (\zeta+s ,p_{K + 1} ) \neq 1$ and
$\gcd(s,N')=1$ implies the second condition $\gcd (\zeta ,p_{K +
1} ) = 1 \neq p_{K + 1} $. Thus, from (\ref{eq13}), $|G_{N',y}
(R_{N',s} )| = p_{K + 1} |G_{N,x} (R_{N,s} )| - 2|G_{N,x} (R_{N,s}
)| = (p_{K + 1}  - 2)\prod\nolimits_{i \in M_x^K } {(p_i  - 2)}  =
\prod\nolimits_{i \in M_y^{K + 1} } {(p_i - 2)}$ when $y=x$ and
$|G_{N',y} (R_{N',s} )| = p_{K + 1} |G_{N,x} (R_{N,s} )| - (p_{K +
1}  - 1)|G_{N,x} (R_{N,s} )| = \prod\nolimits_{i \in M_x^K } {(p_i
- 2)}  = \prod\nolimits_{i \in M_y^{K + 1} } {(p_i  - 2)}$ when
$y=xp_{K+1}$, which concludes the proof.
\end{IEEEproof}
\vskip 24pt
 \indent \textit{\textbf{Lemma }10}\textit{:}
For $N=p_1^{c_1}$, its divisor $x=p_1^{n_1(x)}$, $0< n_1(x)\leq c_1$
and $s \in \mu_N$, $|G_{N,x}(R_{N,s})|$ is given as
\begin{equation}\label{eq14}
\begin{split}
|G_{N,x}(R_{N,s})| = \left\{ {\begin{array}{*{20}c}
   &\phi_x(1),  &&\text{if}\, x=p_1^{n_1(x)}, \\
   &\Phi_x(1),  &&\text{if}\, x=1, \\
   &0, &&\text{otherwise}.\\
\end{array}} \right.\\\\
\end{split}
\end{equation}
\begin{IEEEproof}
When $n_1 (x)=c_1$, $G_{N,N} (R_{N,s} ) = \{\, r - s\,|\,\gcd (r -
s,N) = N,\,r \in \mu _N \,\}  = \{\, 0\,\}$, which implies
$|G_{N,N} (R_{N,s} )| = 1$. When $0 < n_1 (x)<c_1$, $G_{N,p_1^{n_1
(x)} } (R_{N,s} ) = \{ \,r - s\,|\,\gcd (r - s,N) = p_1^{n_1 (x)}
,\,r \in \mu _N \,\}=\{ \,r - s\,|\,r = mp_1^{n_1 (x)} + s,\,0
\leq m < p_1^{c_1 - n_1 (x)} ,\,\gcd (m,p_1 ) = 1\,\}$. Since
there are $p_1^{c_1-n_1(x)-1}$ integer multiples of $p_1$ among
$\{\,m\,|\,0\leq m< p_1^{c_1-n_1(x)}$\}$, |G_{N,p_1^{n_1 (x)}}
(R_{N,s} )| = p_1^{c_1 - n_1 (x)}  - p_1^{c_1  - n_1 (x) - 1}$.
When $x=1$, $|G_{N,1} (R_{N,s} )| = \{ \,r - s\,|\,\gcd (r - s,N)
= 1,\,r \in \mu _N \,\}$. Note that $|G_{N,1} (R_{N,s} )| =
|R_{N,s} | - \sum\nolimits_{x =2}^{N} {|G_{N,x} (R_{N,s} )|} =
|\mu _N | - \sum\nolimits_{i = 1}^{c_1 } {|G_{N,p_1^i } (R_{N,s}
)|}$. Since $|\mu_N |=p_1^{c_1-1} (p_1-1)$ from Lemma 6 and
$|G_{N,N}(R_{N,s})|=1$, $|G_{N,1} (R_{N,s} )| = p_1^{c_1 } -
p_1^{c_1 - 1} - \sum\nolimits_{i = 1}^{c_1  - 1} {(p_1^{c_1 - i} -
p_1^{c_1 - i - 1} } ) - 1 = p_1^{c_1 }  - 2p_1^{c_1  - 1}$.
\end{IEEEproof}
\vskip 24pt \indent Now, the proof of Theorem \ref{theorem3} is
given as follows.
\begin{IEEEproof}[\text{Proof of Theorem \ref{theorem3}}]
When $k=1$, (\ref{eq33}) holds from Lemma 10. Suppose that
(\ref{eq33}) holds for any divisor $x$ of $N$ when $k=K$ and let
$N'=Np_{K+1}^{c_{K+1}}$. Then, the set of all divisors of $N'$ is
given as a $\{\, xp_{K+1}^{l} \,|\, x\in \{\text{all divisors of }
N\},\, 0 \leq l\leq c_{K+1}\,\}$. Note that
$\mu_{N'}=\{\,n\,|\,0<n<N',\,\gcd(n,N')=1\}=\{\,n+mN\,|\,n\in
\mu_N,\, 0\leq m< p_{K+1}^{c_{K+1}},\, \gcd(n+mN,p_{K+1})=1\}$.
Thus, $R_{N',s}=\{\,n-s\,|\,n\in \mu_{N'}\,\}= \mathop  \bigcup
\nolimits_{m \in \{0, \cdots,p_{K+1}^{c_{K+1}} - 1\}} R_{N,s}^m$,
where $R_{N,s}^m = \{\, n + mN - s\,|\,n
\in \mu _N ,\,\gcd (n + mN,p_{K + 1} ) = 1\,\}$.\\
\indent When $y=x$ for a divisor $x$ of $N$, $G_{N',y} (R_{N,s}^m
) = \{ \,n + mN\,|\,n \in G_{N,x} (R_{N,s}),\,\gcd (n + mN,p_{K +
1} ) = 1,\,\gcd (n - s + mN,p_{K + 1} ) = 1\,\}$ since
$\gcd(n-s+mN, N')=x$ implies $\gcd(n-s,N)=x$ and
$\gcd(n-s+mN,N')=x$ if $\gcd(n-s,N)=x$ and
$\gcd(n-s+mN,p_{K+1})=1$. Also, when $y=xp_{K+1}^{l}$ for a
divisor $x$ of $N$ and $1 \leq l \leq c_{K+1}$, $G_{N',y}
(R_{N,s}^m ) = \{ \,n + mN\,|\,n \in G_{N,x} (R_{N,s} ),\,\gcd (n
+ mN,p_{K + 1} ) = 1,\,\gcd (n - s +
mN,p_{K + 1}^{l} ) = p_{K + 1}^{l} \,\}$. \\
\indent Now, define $B_{N} (n) = \{ \,n + mN\,|\,0 \leq m < p_{K +
1}^{c_{K+1}} \,\}$. Then, it is easily seen that
\begin{equation}\label{eq15}
\begin{split}
  |G_{N',y} (R_{N',s})| =& \sum\limits_{m = 0}^{p_{K + 1}^{c_{K+1}}  - 1} {|G_{N',y} (R_{N',s}^m )|}  \hfill \\
   =& \sum\limits_{n \in G_{N,x} (R_{N,s} )}{|B_{N} (n)| - } |\alpha _{N,N'}^s (x,y)|\hfill,
\end{split}
\end{equation}
where $\alpha _{N,N'}^s(x,y) = \{\,\zeta \,|\, \zeta\in \mathop
\bigcup \nolimits_{n \in G_{N,x} (R_{N,s}) } B_{N}(n),$ $\gcd
(\zeta+s ,p_{K + 1} ) \ne 1\,\,\text{or}\,\,\gcd (\zeta ,p_{K +
1}^{c_{K+1}} ) \ne y/x\,\}$. Similarly to the proof of Lemma 9,
$|\alpha _{N,N'}^s (x,y) \cap B_{N} (n)| = 2p_{K + 1}^{c_{K+1}-1}$
when $y=x$ and $|G_{N',y} (R_{N',s})|$ then is given as
\begin{equation}\label{eq16}
\begin{split}
|G_{N',y} (R_{N',s})|&=\!p_{K + 1}^{c_{K+1}}\prod\limits_{i\in
M_{x}} {\phi_x (i)\prod\limits_{j \in M_{x}^K}{\Phi_x (j)}}-2p_{K
+ 1}^{c_{K+1}-1}{ {\prod\limits_{i\in M_{x}}{\phi_x
(i)\prod\limits_{j \in M_{x}^K}{\Phi_x (j)}}}}\\
 &= {
{\prod\limits_{i\in M_{y}}{\phi_y (i)\prod\limits_{j \in
M_{y}^{K+1}}{\Phi_y (j)}}}},
\end{split}
\end{equation}
since $M_y=M_x$, $M_y^{K+1}=M_x^K \cup \{K+1\}$, $\phi_y(i)= \phi_x
(i)$ for $i\in M_x$, $\Phi_y(j)= \Phi_x (j)$ for $j\in M_x^K$ and
$\Phi_y(K+1)=p^{c_{K+1}}_{K+1}-2p^{c_{K+1}-1}_{K+1}$ so that $ \prod
\nolimits_{i \in M_y^{K+1} } \Phi {}_y(j)=\left( {p_{K + 1}^{c_{K +
1}}  - 2p_{K + 1}^{c_{K + 1} - 1} } \right)\mathop \prod
\nolimits_{i \in M_x } \Phi {}_x(i)$. Also, similarly to the proof
of Lemma 9, $|\alpha _{N,N'}^s (x,y) \cap B_{N} (n)|
=p_{K+1}^{c_{K+1}}-1$ when $y=xp_{K+1}^{c_{K+1}}$ and $|G_{N',y}
(R_{N',s})|$ is then given as
\begin{equation}\label{eq17}
\begin{split}
|G_{N',y} (R_{N',s})| &=p_{K + 1}^{c_{K+1}} \prod\limits_{i\in
M_{x}}{\phi_x (i)\prod\limits_{j \in M_{x}^K}\!\!\!\!{\Phi_x
(j)}}-\left( {p_{K + 1}^{c_{K+1}}-1} \right)\!\!\!{
{\prod\limits_{i\in M_{x}}{\phi_x
(i)\prod\limits_{j \in M_{x}^K}{\Phi_x (j)}}}}\\
 &= {
{\prod\limits_{i\in M_{x}}{\phi_x (i)\prod\limits_{j \in
M_{x}^K}{\Phi_x (j)}}}}= { {\prod\limits_{i\in M_{y}}{\phi_y
(i)\prod\limits_{j \in M_{y}^{K+1}}{\Phi_y (j)}}}},
\end{split}
\end{equation}
since $M_y=M_x \cup \{K+1\}$, $M_y^{K+1}=M_x^K$, $\phi_y(i)= \phi_x
(i)$ for $i\in M_x$ and $\phi_y(K+1)=1$ so that ${\mathop \prod
\nolimits_{i \in M_y } \phi_y(i) = \mathop \prod \nolimits_{i \in
M_x } \phi_x(i)}$ and $\mathop \prod \nolimits_{i \in M_y^{K+1} }
\Phi {}_y(j) = \mathop \prod \nolimits_{i \in M_x^{K} } \Phi
{}_x(j)$ for $j\in M_x^K=M_y^{K+1}$. When $y=xp_{K + 1}^{l}$ for
$0<l<{c_{K+1}}$, $ G_{N',y} (R_{N',s} ) = \bigcup\nolimits_{m =
0}^{p_{K + 1}^{c_{K + 1}  - 1} } {G_{N',y} (R_{N',s}^m )}  = \{
\,\zeta \,|\,\zeta \in \bigcup\nolimits_{n \in G_{N,x} (R_{N,s}
)}^{} {B_{N,y} (n),} \,\zeta  = mp_{K + 1}^l + s,\,0 \leq m < p_{K +
1}^{c_{K + 1}  - l} ,\,\gcd (m,p_{K + 1} ) = 1\}$. Similar to the
proof of Lemma 10, $|G_{N',y} (R_{N',s} )|$ is given as
\begin{equation}\label{eq18}
\begin{split}
  \left| {G_{N',y} (R_{N',s} )} \right| =& \left( {p_{K + 1}^{c_{K + 1}  - l}  - p_{K + 1}^{c_{K + 1}  - l - 1} } \right)\left| {G_{N,x} (R_{N,s} )} \right|\hfill \\
   =& \left( {p_{K + 1}^{c_{K + 1}  - l}  - p_{K + 1}^{c_{K + 1}  - l - 1} } \right)\mathop \prod \limits_{i \in M_x } \phi_x(i)\mathop \prod \limits_{j \in M_x^K } \Phi _x(j)\\
   = &\mathop \prod \limits_{i \in M_y } \phi {}_y(i)\mathop \prod \limits_{j \in M_y^{K + 1} } \Phi {}_y(j), \hfill \\
\end{split}
\end{equation}
since $M_y=M_x \cup \{K+1\}$, $M_y^{K+1}=M_x^K$,
$\phi_y(i)=\phi_x(i)$ for $i\in M_x$ and
$\phi_y(K+1)=p_{K+1}^{c_{K+1}-l}-p_{K+1}^{c_{K+1}-l-1}$ so that $
\prod \nolimits_{i \in M_y } \phi {}_y(i)=\left( {p_{K + 1}^{c_{K
+ 1}  - l}  - p_{K + 1}^{c_{K + 1}  - l - 1} } \right)\mathop
\prod \nolimits_{i \in M_x } \phi {}_x(i)$ and $\Phi {}_y(j) =
\Phi {}_x(j)$ for $j \in M_x^K  = M_y^{K + 1},$ which concludes
the proof.
\end{IEEEproof}

\subsection{Number of Available Chu sequences}
In this section, the number of available Chu sequences satisfying
a given maximum magnitude of the cross-correlation is
investigated. The main result of this subsection is given in the
following theorems. \\
\begin{theorem}\label{theorem4}
Let a partial Chu sequence set $C_A$ be defined as
$C_A=\{a_r\,|\,r\in A \,\}$ for a given partial index set $A
\subset \mu_N$ and $\hat \theta _c^A$ be the maximum magnitude of
the cross-correlation among sequences in $C_A$. Then, $\hat \theta
_c^A \leq |\theta|$ if and only if any two elements in
$C_A$, $a_r$ and $a_s$, satisfy $\gcd(r-s, N)\leq |\theta|^2/N$.\\
\end{theorem}
\begin{IEEEproof}
The proof is apparent since $\hat \theta _{r,s}  = \sqrt {N\gcd (r
- s,N)}$ from Theorem \ref{theorem1}.
\end{IEEEproof}
\vskip 24pt
\begin{theorem}
For a given $\sqrt{N} \leq |\theta| \leq N$, let $ X_{N,\theta}=
\left\{ { } \right.$ all divisors of
$N=\prod\nolimits_{i=1}^{k}{p_i^{c_i } }$ greater than $ \left.
{|\theta| ^2/{N}}\, \right\}$, $x_{\min }  = \min X_{N,\theta }$,
$M_{x_{\min}} = \{n | 1 \le n < x_{\min}, \gcd(n,N)=1 \}$, and $
x_{\varphi _{\min } }  = \mathop {\arg \,\min }\nolimits_{x \in
X_{N,\theta } }  {\varphi \left( {x } \right)}$. Then, the largest
cardinality among partial Chu sequence sets satisfying $\hat \theta
_c^A \leq |\theta|$ is lower bounded by $|M_{x_{\min}}|$ and upper
bounded by $\varphi(x_{\varphi_{\min } })$. \\
\label{theorem5}
\end{theorem}
\begin{IEEEproof}
The lower bound is apparent from Theorem \ref{theorem4} since the
difference of any two elements in $M_{x_{\min}}$ is smaller than
$x_{\min}$. Now, consider the upper bound part. Define
\begin{equation}\label{eq19}
R_{x_{\varphi _{\min } } } = \cup_{n \in N_{x_{\varphi_{\min } }}}
R_{x_{\varphi _{\min } } }^n,
\end{equation}
where $R_{x_{\varphi _{\min } } }^n = \{m x_{\varphi_{\min } }+n | 0
\le m < N/x_{\varphi_{\min } } \}$ and $N_{x_{\varphi_{\min }
}}=\{n| 1 \le n < x_{\varphi_{\min } }, \gcd(n,x_{\varphi_{\min }
})=1 \}$. Then, it is apparent that $\mu_n \subset R_{x_{\varphi
_{\min } } }$ and $|N_{x_{\varphi_{\min }
}}|=\varphi(x_{\varphi_{\min } })$. Let $A$ be any partial Chu
sequence set satisfying $\hat \theta _c^A \leq |\theta|$. Then, $A
\cap R_{x_{\varphi _{\min } } }^n \le 1$ because the difference of
any two distinct elements in $R_{x_{\varphi _{\min } } }^n$ is at
least $x_{\min}$. Thus, we can pick at most one element in
$R_{x_{\varphi _{\min } } }^n$ for each $n \in N_{x_{\varphi_{\min }
}}$, which proves the upper bound.
\end{IEEEproof}

\vskip 24pt

\textit{\textbf{Lemma }11}\textit{:} $R_{x_{\varphi _{\min } } }^n
\cap \mu_n \neq \varnothing$ for any $n \in N_{x_{\varphi_{\min }
}}$. \\

\begin{IEEEproof}
Since $\gcd(n,x_{\varphi_{\min } })=1$, $\gcd(mx_{\varphi_{\min }
}+n,N)=\gcd(n, (N/x_{\varphi_{\min } }-m)x_{\varphi_{\min } })=
\gcd(n, N/x_{\varphi_{\min } }$$\\-m)$. If $\gcd(n,N)=1$,
$\gcd(mx_{\varphi_{\min } }+n,N)=1$ when $m=0$. If
$\gcd(n,N)=g>1$, $\gcd(mx_{\varphi_{\min } }+n,N)=1$ when $1 \le
m={N(g-1)}/{(x_{\varphi_{\min } }g)}<N/x_{\varphi_{\min } }$,
which concludes the proof.
\end{IEEEproof}

Note that Lemma 11 implies that the upper bound in Theorem
\ref{theorem5} is tight. Also, although not shown explicitly,
exhaustive search showed that we can find a partial Chu sequence
set with cardinality equal to the upper bound for the sequence
length $N$ up to $10^5$. Thus, although not proved, we may
conjecture that the largest cardinality among partial Chu sequence
sets satisfying $\hat \theta _c^A \leq |\theta|$ is
$\varphi(x_{\varphi_{\min } })$.

\textit{\text{Example }1} ($N=143$, $|\theta|=\sqrt{1430}$) : In
this example, $x_{\min}={x_{\varphi _{\min } } }=11$,
$R_{x_{\varphi _{\min } } }=\{11m + n\,\left| \,0 \right.$ $  \le
m < 13 \,\,\text{and}\,\,\,1 \le n < 11\}$  and $\varphi \left(
{x_{\varphi _{\min } } } \right)=$ $10$. It is easily found that
$C_A$, $A=\left\{ {1,2, \cdots ,10} \right\}$, satisfies $\hat
\theta _c^A \leq |\theta|$ and $|A|=\varphi \left( {x_{\varphi
_{\min } } } \right)$. \\

\indent \textit{\text{Example }2} ($N=154$,
$|\theta|=\sqrt{1540}$) : In this example, $x_{\min}=11$,
${x_{\varphi _{\min } } }=14$, $R_{x_{\varphi _{\min } } }  = \{
14m + n\,\left| \,0 \right. $ $ \le m < 11, 1 \le n < 14,
\gcd(n,14)=1 \}$ and $\varphi \left( {x_{\varphi _{\min } } }
\right)=6$. Note that $\left\{ n | 1 \le n < 14, \gcd(n,14)=1
\right\} = \{ 1, 3, 5, 9, 11, 13 \}$ and $11$ is not relatively
prime with $N$ so that it cannot be included in a partial Chu
sequence set. Instead of $11$, pick $14 \times 2 + 11 = 39$ to
construct $A=\{ 1, 3, 5, 9, 13, 39 \}$. Then, it is easily found
that $C_A$ satisfies $\hat \theta _c^A \leq |\theta|$ and
$|A|=\varphi \left( {x_{\varphi
_{\min } } } \right)$. \\

The three subfigures in Fig.~1 shows the number of available Chu
sequences satisfying $|\hat \theta _{A}/N|^2<\theta_N^2$ for
$\theta_N^2=0.01, 0.05, 0.1$, when $N$ is around 512, 1024, and
2048, respectively. Also, Table I gives the prime factors of $N$
used in Fig. 1. It is well known that $N$ is preferred to be a
prime number, which is confirmed from the results. Also, the
results indicate that, in cases we must choose $N$ among non-prime
numbers by some reason, the number of available Chu sequences
tends to increase as the smallest difference between prime factors
increases. Thus, it is preferred to choose $N$ composed of two
prime factors with relatively large difference (e.g., 508, 514,
515, 2045, 2049, 2051 as shown in Table I.

\section{Conclusions}
In this paper, we analyzed generalized cross-correlation properties
for Chu sequences. From the analysis, it was obtained that i) the
magnitude of the cross-correlation function between any two Chu
sequences for all possible lags, ii) the distribution of the maximum
magnitude of the cross-correlation among a given Chu sequence set
and iii) the number of available Chu sequences satisfying a given
cross-correlation constraint.
\appendix
\subsection{Proof of Lemma 1}
When $N$ is an even number, we can rewrite $\theta_{r,s} (\tau)$
as
\begin{equation}\label{eq20}
\begin{split}
\theta _{r,s} \left( \tau  \right)&= \sum\limits_{k = 0}^{N - 1}
{\exp \left( {j\pi \frac{{rk^2 }} {N}} \right)} \exp \left( { -
j\pi \frac{{s\left( {k + \tau } \right)^2 }}
{N}} \right) \hfill \\
&= \exp \left( { - \pi \frac{{s\tau ^2 }} {{u_{r,s} g_{r,s} }}}
\right) \sum\limits_{k = 0}^{u_{r,s} g_{r,s} - 1}\!\!{\exp \left\{
{j2\pi \left( {\frac{{v_{r,s} k^2 }} {{2u_{r,s} }} -
\frac{{sk\left( {i_\tau  g_{r,s}  + d_\tau } \right)}} {{u_{r,s}
g_{r,s} }}} \right)} \right\}}.  \hfill
\end{split}
\end{equation}
Then, the squared magnitude, $\left| {\theta_{r,s} (\tau)}
\right|^2$, is given as
\begin{equation}\label{eq21}
\begin{split}
\left| {\theta _{r,s} \left( \tau  \right)} \right|^2 & =  \hfill \\
&\!\!\!\!\sum\limits_{k = 0}^{u_{r,s} g_{r,s}  - 1}
{\sum\limits_{l = 0}^{u_{r,s} g_{r,s}  - 1} {\exp \left\{ {j2\pi
\left( {\frac{{v_{r,s} k^2 }} {{2u_{r,s} }} - \frac{{sk\left(
{i_\tau g_{r,s}  + d_\tau  } \right)}} {{u_{r,s} g_{r,s} }}}
\right)} \right\}} }  \hfill  \exp \left\{ {j2\pi \left(
{\frac{{sl\left( {i_\tau g_{r,s} + d_\tau  } \right)}} {{u_{r,s}
g_{r,s} }} - \frac{{v_{r,s} l^2 }} {{2u_{r,s} }}} \right)}
\right\}. \hfill
\end{split}
\end{equation}
The last term in (\ref{eq21}) is periodic with period
$u_{r,s}g_{r,s}$ because
\begin{equation}\label{eq22}
\begin{split}
 &\exp \left\{ {j2\pi\ \left({\frac{{s(l + u_{r,s} g_{r,s})(i_\tau g_{r,s} + d_\tau)}}{{u_{r,s} g_{r,s}}} - \frac{{v_{r,s} \left( {l + u_{r,s} g_{r,s}} \right)^2 }}{{2u_{r,s} }}} \right)}\right\}\\
  &= \exp \left\{ {j2\pi \left( {\frac{{sl(i_\tau g_{r,s} + d_\tau)}}{{u_{r,s} g_{r,s}}} - \frac{{v_{r,s} l^2 }}{{2u_{r,s} }}} \right )} \right \}\exp \left( { - j\pi  {{v_{r,s} u_{r,s} g_{r,s}^2 }}} \right)\exp \left\{ {j2\pi (i_\tau g_{r,s} + d_\tau + lg_{r,s}v_{r,s} )} \right\} \\
  &=\exp \left\{ {j2\pi \left( {\frac{{sl(i_\tau g_{r,s} + d_\tau)}}{{u_{r,s} g_{r,s}}} - \frac{{v_{r,s} l^2 }}{{2v_{r,s} }}} \right)}\right\},
\end{split}
\end{equation}
where the last equality comes from the fact that $g_{r,s}$ is
always even so that $\exp \left( { - j\pi v_{r,s} u_{r,s}
g_{r,s}^2 } \right) = 1$ because $r-s$ should be even when $N$ is
even. For a periodic function $f(x)$ with period $N$, it is easily
seen that
\begin{equation}\label{eq23}
\sum\limits_{l = 0}^{N  - 1} {f(l)} =\sum\limits_{e = k}^{N+k - 1
} {f(e)}.
\end{equation}
Then, from (\ref{eq22}) and (\ref{eq23}), (\ref{eq21}) can be
rewritten as
\begin{equation}\label{eq24}
\begin{split}
\left| {\theta _{r,s} \left( \tau  \right)} \right|^2
=\sum\limits_{k = 0}^{u_{r,s} g_{r,s}  - 1}& {\exp \left\{ {j2\pi
\left( {\frac{{v_{r,s} k^2 }} {{2u_{r,s} }} - \frac{{sk\left(
{i_\tau  g_{r,s}  + d_\tau  } \right)}}
{{u_{r,s} g_{r,s} }}} \right)} \right\}}  \hfill \cdot \\
\sum\limits_{e = k}^{u_{r,s} g_{r,s}  + k - 1}\!\!\!&{\exp \left\{
{j2\pi \left( {\frac{{se\left( {i_\tau g_{r,s} + d_\tau }
\right)}} {{u_{r,s} g_{r,s} }} - \frac{{v_{r,s} e^2}}
{{2u_{r,s} }}} \right)} \right\}}  \hfill \\
=\sum\limits_{k = 0}^{u_{r,s} g_{r,s}  - 1} &{\exp \left\{ {j2\pi
\left( {\frac{{v_{r,s} k^2 }} {{2u_{r,s} }} - \frac{{sk\left(
{i_\tau  g_{r,s}  + d_\tau  } \right)}}
{{u_{r,s} g_{r,s} }}} \right)} \right\}}  \hfill \cdot  \\
\sum\limits_{e = 0}^{u_{r,s} g_{r,s} - 1} &{\exp \left\{ {j2\pi \!
\left( {\frac{{s(e+k)\left( {i_\tau g_{r,s}  + d_\tau } \right)}}
{{u_{r,s} g_{r,s} }} \!- \!\frac{{v_{r,s} (e+k)^2}}
{{2u_{r,s} }}} \right)} \right\}}  \hfill \\
= \sum\limits_{e = 0}^{u_{r,s} g_{r,s}  - 1} &{\exp \left\{ {j2\pi
\left( {\frac{{v_{r,s} e^2 }} {{2u_{r,s} }} - \frac{{se\left(
{i_\tau  g_{r,s}  + d_\tau  } \right)}} {{u_{r,s} g_{r,s} }}}
\right)} \right\}}  \hfill \sum\limits_{k = 0}^{u_{r,s} g_{r,s} -
1} {\exp \left( { - j2\pi \frac{{v_{r,s} ke}}{{u_{r,s} }}}
\right)} \hfill.
\end{split}
\end{equation}
The last term can be divided in two terms in (\ref{eq24}), when
$e=mu_{r,s}$ and $e \ne mu_{r,s}$ for $0\leq m<g_{r,s}$.
Therefore, it can be expressed as
\begin{equation}\label{eq25}
\begin{split}
\left| {\theta _{r,s} \left( \tau  \right)} \right|^2
=\sum\limits_{e = mu_{r,s} }&{\exp \left\{ {j2\pi \left(
{\frac{{v_{r,s} e^2 }} {{2u_{r,s} }} - \frac{{se\left( {i_\tau
g_{r,s}  + d_\tau  }\right)}} {{u_{r,s} g_{r,s} }}} \right)}
\right\}}  \hfill \sum\limits_{k = 0}^{u_{r,s} g_{r,s}  - 1} {\exp
\left( { - j2\pi \frac{{v_{r,s} ke}}
{{u_{r,s} }}} \right)}  \hfill \\
+ \sum\limits_{e \ne mu_{r,s} }&{\exp \left\{ {j2\pi \left(
{\frac{{v_{r,s} e^2 }} {{2u_{r,s} }} - \frac{{se\left( {i_\tau
g_{r,s}  + d_\tau  } \right)}} {{u_{r,s} g_{r,s} }}} \right)}
\right\}}  \hfill  \sum\limits_{k = 0}^{u_{r,s} g_{r,s} - 1} {\exp
\left( { - j2\pi \frac{{v_{r,s} ke}} {{u_{r,s} }}} \right)}.
\hfill
\end{split}
\end{equation}
When $e \ne mu_{r,s}$ for $0\leq m<g_{r,s}$, the last term is
equal to 0 from Lemma 1 in (\ref{eq25}) because $u_{r,s}$ is
relatively prime with $v_{r,s}$. Accordingly, we can rewrite
(\ref{eq25}) as follows
\begin{equation}\label{eq26}
\begin{split}
\left| {\theta _{r,s} \left( \tau  \right)} \right|^2
&=\sum\limits_{e = mu_{r,s} }\!\!\!{\exp \left\{ {j2\pi \left(
{\frac{{v_{r,s} e^2 }} {{2u_{r,s} }} - \frac{{se\left( {i_\tau
g_{r,s}  + d_\tau  } \right)}}{{u_{r,s} g_{r,s} }}} \right)}
\right\}}  \hfill \sum\limits_{k = 0}^{u_{r,s} g_{r,s}  - 1} {\exp
\left( { -j2\pi \frac{{v_{r,s} ke}}{{u_{r,s} }}} \right)}  \hfill \\
&=\sum\limits_{m = 0}^{g_{r,s}  - 1}\! {\exp \!\left\{ {j2\pi
\left( {\frac{{v_{r,s} u_{r,s} m^2 }} {2}\! -\!\! \frac{{sm\left(
{i_\tau g_{r,s}  + d_\tau  } \right)}}{{g_{r,s} }}} \right)}
\right\}}  \hfill \sum\limits_{e = 0}^{u_{r,s} g_{r,s}  - 1} {\exp \left( { - j2\pi v_{r,s} me} \right)}  \hfill \\
&= u_{r,s} g_{r,s}  \sum\limits_{m = 0}^{g_{r,s}  - 1} {\left( { -
1} \right)^{v_{r,s} u_{r,s} m^2 } \exp \left( { {\frac{{smd_\tau
}} {{g_{r,s} }}} } \right)}.  \hfill
\end{split}
\end{equation}
When $N$ is an odd number, we can derive $\left| {\theta_{r,s}
(\tau)} \right|^2$ in similar to case when $N$ is even.
Accordingly, we can rewrite $\left| {\theta_{r,s} (\tau)}
\right|^2$ as
\begin{equation}\label{eq27}
\begin{split}
\theta _{r,s} \left( \tau  \right)&= \sum\limits_{k = 0}^{N - 1}
{\exp \left( {j\pi \frac{{rk\left( {k + 1} \right)}} {N}} \right)}
\exp \left( { - j\pi \frac{{s\left( {k + \tau } \right)\left( {k +
1 + \tau } \right)}}
{N}} \right) \hfill \\
&=\exp \left( { - j\pi \frac{{s\left( {\tau ^2  + \tau } \right)}}
{{u_{r,s} g_{r,s} }}} \right)\sum\limits_{k = 0}^{u_{r,s} g_{r,s}
- 1} {\exp \left\{ {j2\pi \left( {\frac{{v_{r,s} \left( {k^2  + k}
\right)}} {{2u_{r,s} }} - \frac{{sk\left( {i_\tau g_{r,s}  +
d_\tau  } \right)}}
{{u_{r,s} g_{r,s} }}} \right)} \right\}}.  \hfill \\
\end{split}
\end{equation}
Then, the squared magnitude, $\left| {\theta_{r,s} (\tau)}
\right|^2$, is given as
\begin{equation}\label{eq28}
\begin{split}
&\left| {\theta _{r,s} \left( \tau  \right)} \right|^2  =\\
&\sum\limits_{k = 0}^{u_{r,s} g_{r,s}  - 1}  {\sum\limits_{l =
0}^{u_{r,s} g_{r,s}  - 1}   \! \! \! \! \! \!{\exp   \!\left\{ \!
{j2\pi   \left( {\frac{{v_{r,s} \left( {k^2 + k} \right)}}
{{2u_{r,s} }}  -  \frac{{sk\left( {i_\tau g_{r,s}  + d_\tau  }
\right) }} {{u_{r,s} g_{r,s} }}} \right)}  \! \!\right\}} } \hfill
\exp \left\{ {j2\pi \left( {\frac{{sl\left( {i_\tau g_{r,s} +
d_\tau  } \right)}} {{u_{r,s} g_{r,s} }} - \frac{{v_{r,s} \left(
{l^2  + l} \right)}}
{{2u_{r,s} }}} \right)} \right\}. \hfill \\
\end{split}
\end{equation}
The last term in (\ref{eq28}) is periodic with period
$u_{r,s}g_{r,s}$ because
\begin{equation}\label{eq29}
\begin{split}
&\exp \left\{ {j2\pi \frac{{s\left( {l + u_{r,s} g_{r,s} }
\right)\left( {i_\tau  g_{r,s}  + d_\tau  } \right)}} {{u_{r,s}
g_{r,s} }}} \right\} \exp \left\{ { - j\pi \frac{{v_{r,s} \left(
{\left( {l + u_{r,s} g_{r,s} } \right)^2  + \left( {l + u_{r,s}
g_{r,s} } \right)} \right)}}
{{u_{r,s} }}} \right\} \hfill \\
&= \exp \!\left\{ {j2\pi\! \left( {\frac{{sl\left( {i_\tau g_{r,s}
+ d_\tau  } \right)}} {{u_{r,s} g_{r,s} }} - \frac{{v_{r,s} \left(
{l^2  + l} \right)}}
{{2u_{r,s} }}} \right)} \right\} \exp \!\left( {\! - j\pi v_{r,s}g_{r,s}u_{r,s}\!\left( {g_{r,s}\!  +\! 1\!} \right)} \right)\!\exp\! \left( {j2\pi \!\left( {i_\tau  g_{r,s} \! + \!d_\tau \!  +\! lg _{r,s} v_{r,s} } \right)} \right) \hfill \\
&=\exp \left\{ {j2\pi \left( {\frac{{sl\left( {i_\tau  g_{r,s}  +
d_\tau  } \right)}} {{u_{r,s} g_{r,s} }} - \frac{{v_{r,s} \left(
{l^2  + l} \right)}}
{{2u_{r,s} }}} \right)} \right\},\hfill\\
\end{split}
\end{equation}
where the last equality comes from the fact that $g_{r,s}+1$ is
always even so that $\exp \left( { - j\pi v_{r,s}g_{r,s} \left(
{u_{r,s}g_{r,s} + 1} \right)} \right)= 1$ because $r-s$ should be
odd when $N$ is odd. Then, from (\ref{eq23}) and (\ref{eq29}),
(\ref{eq28}) can be rewritten as
\begin{equation}\label{eq30}
\begin{split}
\left| {\theta _{r,s} \left( \tau  \right)} \right|^2  =
\sum\limits_{k = 0}^{u_{r,s} g_{r,s}  - 1}& {\exp \left\{ {j2\pi
\left( {\frac{{v_{r,s} \left( {k^2  + k} \right)}} {{2u_{r,s} }} -
\frac{{sk\left( {i_\tau  g_{r,s}  + d_\tau  } \right)}}
{{u_{r,s} g_{r,s} }}} \right)} \right\}}\cdot   \hfill \\
\sum\limits_{e = k}^{u_{r,s} g_{r,s}  + k - 1}\ \!\!\!\!\!&{\exp
\left\{ {j2\pi \left( {\frac{{se\left( {i_\tau  g_{r,s}  + d_\tau
} \right)}} {{u_{r,s} g_{r,s} }} - \frac{{v_{r,s} \left( {e^2  +
e} \right)}}
{{2u_{r,s} }}} \right)} \right\}}  \hfill \\
= \sum\limits_{e = 0}^{u_{r,s} g_{r,s}  - 1} &{\exp \left\{ {j2\pi
\left( {\frac{{v_{r,s} \left( {e^2  + e} \right)}} {{2u_{r,s} }} -
\frac{{se\left( {i_\tau  g_{r,s}  + d_\tau  } \right)}} {{u_{r,s}
g_{r,s} }}} \right)} \right\}} \sum\limits_{k = 0}^{u_{r,s}
g_{r,s}  - 1}{\exp \left( { - j2\pi \frac{{v_{r,s} ke}} {{u_{r,s}
}}} \right)}.  \hfill
\end{split}
\end{equation}
The last term can be divided in two terms in (\ref{eq30}), when
$e=mu_{r,s}$ and $e \ne mu_{r,s}$ for $0\leq m<g_{r,s}$.
Therefore, it can be expressed as
\begin{equation}\label{eq31}
\begin{split}
\left| {\theta _{r,s} \left( \tau  \right)} \right|^2
=\sum\limits_{e = mu_{r,s} }&{\exp \left\{ {j2\pi \left(
{\frac{{v_{r,s} (e^2+e) }} {{2u_{r,s} }} -\frac{{se\left( {i_\tau
g_{r,s}  + d_\tau  } \right)}} {{u_{r,s} g_{r,s} }}} \right)}
\right\}}  \sum\limits_{k = 0}^{u_{r,s} g_{r,s}  - 1} {\exp \left(
{ - j2\pi \frac{{v_{r,s} ke}}
{{u_{r,s} }}} \right)}  \hfill \\
+ \sum\limits_{e \ne mu_{r,s} }&{\exp \left\{ {j2\pi \left(
{\frac{{v_{r,s} (e^2+e) }} {{2u_{r,s} }} - \frac{{se\left( {i_\tau
g_{r,s}  + d_\tau  } \right)}} {{u_{r,s} g_{r,s} }}} \right)}
\right\}}  \hfill \sum\limits_{k = 0}^{u_{r,s} g_{r,s} - 1} {\exp
\left( { - j2\pi \frac{{v_{r,s} ke}} {{u_{r,s} }}} \right)}.
\hfill
\end{split}
\end{equation}
When $e \ne mu_{r,s}$ for $0\leq m<g_{r,s}$, the last term is
equal to 0 from Lemma 1 in (\ref{eq31}) because $u_{r,s}$ is
relatively prime with $v_{r,s}$ and we can rewrite (\ref{eq31}) as
follows
\begin{equation}\label{eq32}
\begin{split}
\left| {\theta _{r,s} \left( \tau  \right)} \right|^2
&=\sum\limits_{e = mu_{r,s} }{\exp \left\{ {j2\pi \left(
{\frac{{v_{r,s} (e^2+e) }} {{2u_{r,s} }} - \frac{{se\left( {i_\tau
g_{r,s}  + d_\tau  } \right)}}{{u_{r,s} g_{r,s} }}} \right)}
\right\}}\sum\limits_{k = 0}^{u_{r,s} g_{r,s}  - 1} {\exp
\left( { -j2\pi \frac{{v_{r,s} ke}}{{u_{r,s} }}} \right)}  \hfill \\
&=\sum\limits_{m = 0}^{g_{r,s}  - 1}\! {\exp \left\{ {j2\pi \left(
{\frac{{v_{r,s} m(u_{r,s}  + 1) }} {2} - \frac{{sm\left( {i_\tau
g_{r,s}  + d_\tau  } \right)}}{{g_{r,s}
}}} \right)} \right\}}\sum\limits_{e = 0}^{u_{r,s} g_{r,s}  - 1} {\exp \left( { - j2\pi v_{r,s} me} \right)}  \hfill \\
&= u_{r,s} g_{r,s} \sum\limits_{m = 0}^{g_{r,s}  - 1} {\left( { -
1} \right)^{v_{r,s} m(u_{r,s}  + 1)} {\exp \left( {
{\frac{{smd_\tau }} {{g_{r,s} }}} } \right)}}.  \hfill
\end{split}
\end{equation}
On that way we have proved Lemma 1.

\section*{Acknowledgment}
This research was supported by the MKE (Ministry of Knowledge
Economy), Korea, under the ITRC (Information Technology Research
Center) support program supervised by the IITA (Institute of
Information Technology Assessment) (IITA-2008-(C1090-0801-0011))

\begin{figure}[ht]
\begin{center}\label{f1}

\includegraphics[width=4.8in, height=3.2in]{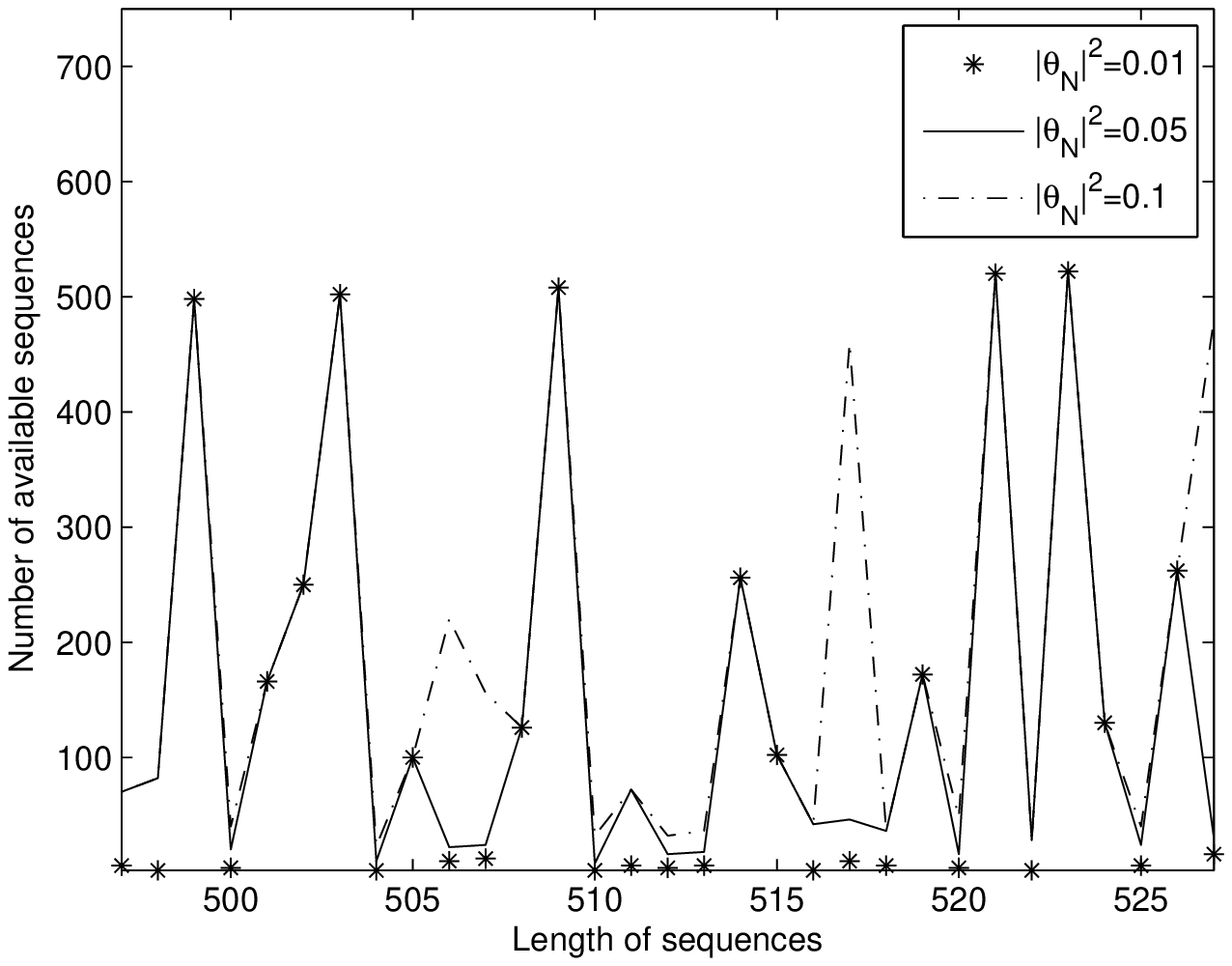}
\includegraphics[width=4.8in, height=3.2in]{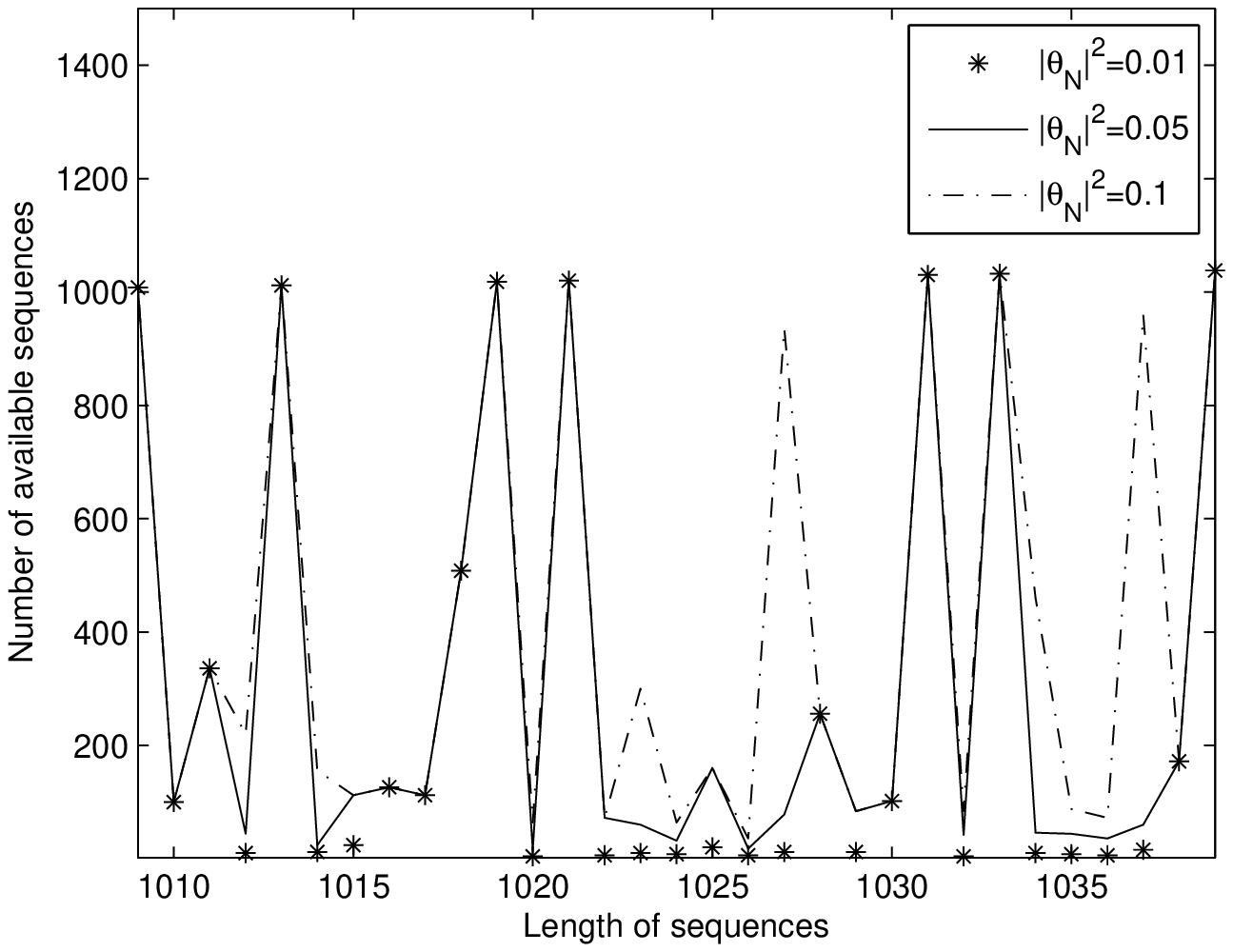}
\includegraphics[width=4.8in, height=3.2in]{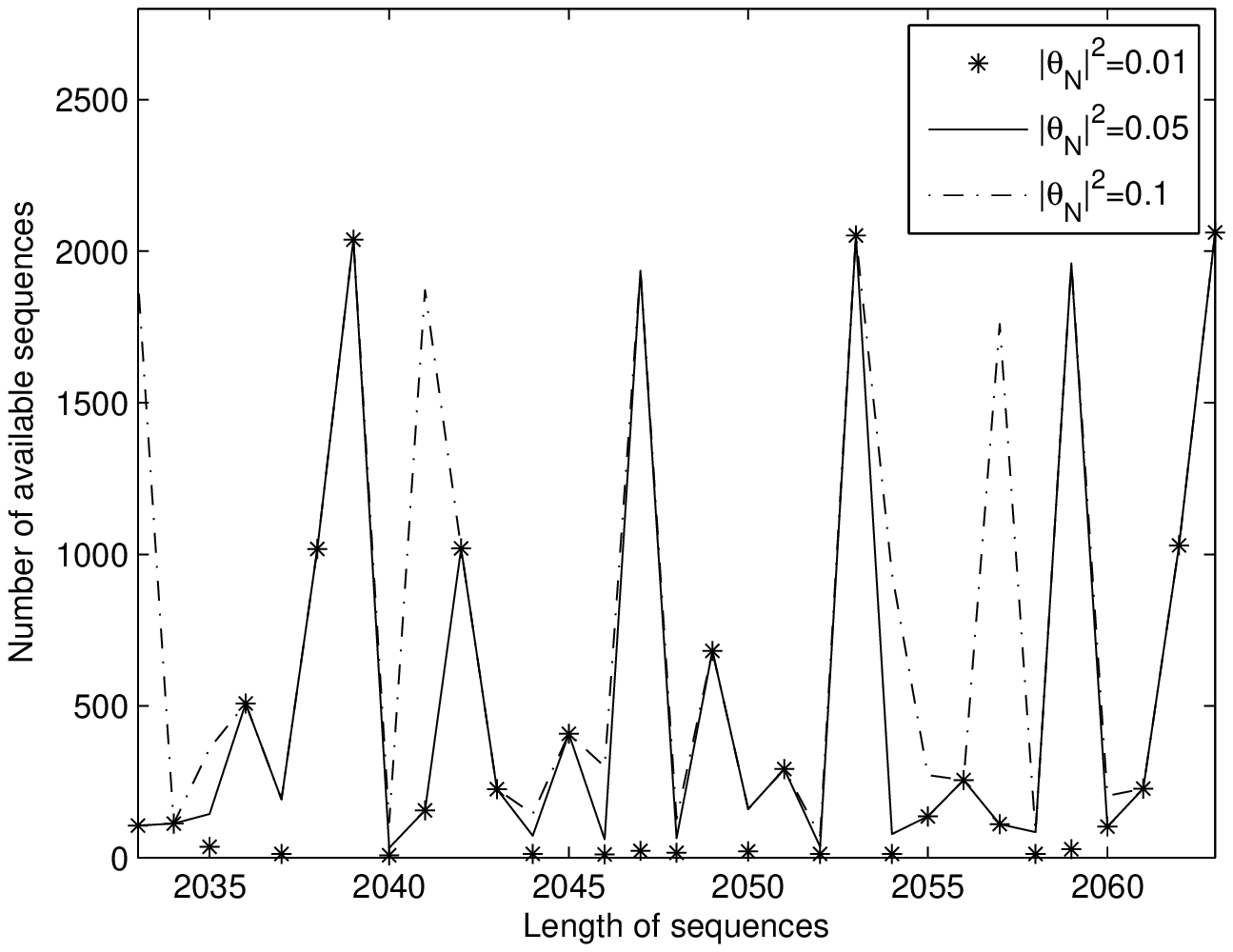}
\caption{Number of available sequences for a given maximum
magnitude bound of the cross-correlation function}
\end{center}
\end{figure}
\newpage
\begin{table*}[ht] \caption{The prime factors of adjacent numbers for $N=512$, $1024$ and $2048$}
\begin{center}
\begin{tabular}{|c||c|c|c|c|c|c|c|c|c|c|c|}
\hline
$N$ & 507&508&509&510&511&512&513&514&515&516&517\\
\hline $\textbf{Prime factors} $&$3,13^2$&  $2^2,127$& $509$&
$2,3,5,17$& $7,73$&
$2^9$& $3^3,19$& $2, 257$&  $5,103$&   $2^2,3,43$&    $11,47$\\
\hline \hline
$N$ & 1019&1020&1021&1022&1023&1024&1025&1026&1027&1028&1029\\
\hline $\textbf{Prime factors}$&$1019$&    $2^2,3,5,17$&  $1021$&
$2,7,73$& $3,11,31$& $2^{10}$& $5^2,41$& $2,3^3,19$&    $13,79$&
$2^2,257$&
$3,7^3$\\
\hline \hline
$N$ & 2043&2044&2045&2046&2047&2048&2049&2050&2051&2052&2053\\
\hline $\textbf{Prime factors}$&$3^2,227$&    $2^2,7,73$& $5,409$&
$2,3,11,31$& $23,89$& $2^{11}$& $3,683$& $2,5^2,41$& $7,293$&
$2^2,3^3,19$& $2053$
\\
\hline
\end{tabular}
\end{center}
\end{table*}

\end{document}